\documentclass[pra,twocolumn,nofootinbib,superscriptaddress]{revtex4-2}
\usepackage{calrsfs}
\usepackage{ulem}
\usepackage{graphics}
\usepackage{epstopdf}
\usepackage{bm}
\usepackage{bbm}
\usepackage{amssymb}
\usepackage{epstopdf}
\usepackage{graphicx}
\usepackage[caption=false]{subfig}
\usepackage{amsmath}
\usepackage{color}
\usepackage{array}

\newcommand{\COR}[1]{{\color{black} #1}}

\begin{document}

\title{Quantum enhanced beam tracking surpassing the Heisenberg uncertainty limit}

\author{Yingwen \surname{Zhang}}
\email{yzhang6@uottawa.ca}
\affiliation{Nexus for Quantum Technologies, University of Ottawa, Ottawa ON Canada, K1N6N5}
\affiliation{National Research Council of Canada, 100 Sussex Drive, Ottawa ON Canada, K1A0R6}
\affiliation{Joint Centre for Extreme Photonics, National Research Council and University of Ottawa, Ottawa, Ontario, Canada}

\author{Duncan \surname{England}}
\affiliation{National Research Council of Canada, 100 Sussex Drive, Ottawa ON Canada, K1A0R6}

\author{Noah \surname{Lupu-Gladstein}}
\affiliation{National Research Council of Canada, 100 Sussex Drive, Ottawa ON Canada, K1A0R6}

\author{Fr\'ed\'eric \surname{Bouchard}}
\affiliation{National Research Council of Canada, 100 Sussex Drive, Ottawa ON Canada, K1A0R6}

\author{Guillaume \surname{Thekkadath}}
\affiliation{National Research Council of Canada, 100 Sussex Drive, Ottawa ON Canada, K1A0R6}

\author{Philip J. \surname{Bustard}}
\affiliation{National Research Council of Canada, 100 Sussex Drive, Ottawa ON Canada, K1A0R6}

\author{Ebrahim \surname{Karimi}}
\affiliation{Nexus for Quantum Technologies, University of Ottawa, Ottawa ON Canada, K1N6N5}
\affiliation{National Research Council of Canada, 100 Sussex Drive, Ottawa ON Canada, K1A0R6}
\affiliation{Joint Centre for Extreme Photonics, National Research Council and University of Ottawa, Ottawa, Ontario, Canada}

\author{Benjamin \surname{Sussman}}
\affiliation{National Research Council of Canada, 100 Sussex Drive, Ottawa ON Canada, K1A0R6}
\affiliation{Nexus for Quantum Technologies, University of Ottawa, Ottawa ON Canada, K1N6N5}
\affiliation{Joint Centre for Extreme Photonics, National Research Council and University of Ottawa, Ottawa, Ontario, Canada}

\begin{abstract} 
Determining a beam's full trajectory requires tracking both its position and momentum (angular) information. However, the product of position and momentum uncertainty in a simultaneous measurement of the two parameters is bound by the Heisenberg uncertainty limit (HUL). In this work, we present a proof-of-principle demonstration of a quantum-enhanced beam tracking technique, leveraging the inherent position and momentum entanglement between photons produced via spontaneous parametric down-conversion (SPDC). We show that quantum entanglement can be exploited to achieve a beam tracking accuracy beyond the HUL in a simultaneous measurement. Moreover, with existing detection technologies, it is already possible to achieve near real-time beam tracking capabilities at the single-photon level. The technique also exhibits high resilience to background influences, with negligible reduction in tracking accuracy even when subjected to a disruptive beam that is significantly brighter than SPDC.
\end{abstract}
\maketitle


Beam tracking plays a crucial role in various technological and scientific applications, ranging from optical communications and remote sensing to high-precision imaging and laser-guided systems. Accurately determining a beam's full trajectory requires tracking both its position and momentum (angular) information. When using a coherent laser beam, the measurement accuracy is fundamentally bounded by the shot noise limit (SNL) when measuring each parameter, and when measuring both parameters simultaneously, the product of the uncertainties is bounded by the Heisenberg uncertainty limit (HUL). 

Many studies have been conducted to beat classical measurement limits by using specific quantum states of light in sensing applications~\cite{Pirandola2018} such as quantum metrology~\cite{Polino2020,Barbieri2022} and quantum imaging~\cite{Moreau2019,Omar2019,Defienne2024}. In the field of beam tracking, spatially squeezed light has been proposed and demonstrated as a method to surpass the SNL~\cite{Fabre2000, Treps2002, Treps2003}. Furthermore, theoretical studies have explored the use of the spatial correlations inherent to entangled photons to beat both the SNL~\cite{Lyons2016} and the HUL~\cite{Lorenzo2013, Bullock2014}. It has also been theoretically shown that the quantum Cram\'er-Rao bound for beam displacement can be achieved using two-photon interference with spatial correlations~\cite{Triggiani2024}.


\COR{For over two decades, it has been known that the product of the position-momentum correlation uncertainty of photons generated via spontaneous parametric down-conversion (SPDC) can surpass the Heisenberg uncertainty limit (HUL)~\cite{Kim1999,Howell2004,Reid2009}, serving as a signature of position-momentum entanglement. Since these initial discoveries, numerous experimental studies have focused on measuring these correlations with progressively higher accuracy and speed, leveraging advancements in detection technologies and techniques~\cite{Edgar2012,Moreau2014,Ndagano2020,Courme2023}. This effect has also been demonstrated in the time-frequency basis~\cite{Maclean2018}.

In this work, we present a practical application of this quantum property by experimentally demonstrating quantum correlation beam tracking (QCBT). Here, position and momentum-entangled photon pairs, generated through SPDC, are used to track a beam's full trajectory—both its position and momentum. The product of the uncertainties in the change in position and momentum serves as an overall measure of the combined accuracy in both planes. We show that this uncertainty product can be measured using QCBT with a precision that exceeds the HUL. Furthermore, we show that QCBT, and thus position-momentum correlation measurements, can be performed in real time at a speed of $\sim1$\,Hz using current detection technologies. This represents an improvement of more than two orders of magnitude in speed compared to previous experiments measuring position-momentum correlations~\cite{Edgar2012,Moreau2014,Ndagano2020,Courme2023}. Additionally, QCBT exhibits exceptional resilience to background noise, owing to the inherent temporal and spatial correlations of SPDC photons.}


To track the full trajectory of a light beam, the change in the beam location will need to be measured simultaneously in two different planes to determine both the position and momentum information. When using a camera-type detector to measure the beam in the position space ${\bm{r}}=(x,y)$ or the momentum space ${\bm{k}}=(u,v)$, the beam location is commonly determined via a centre-of-mass equation
\begin{equation}
    \bar{\bm{r}} = \frac{\sum_{j=1}^{N_s} \bm{r}_j}{N_s},
\label{centroid}
\end{equation}
with $N_s$ the total number of detected photons and $\bm{r}_j$ the position where each photon is detected. This is the most optimal estimator of the beam centroid and the variance of the centroid in position is given by (see the Supplementary Materials for derivation)
\begin{equation}
    \text{Var}(\bar{\bm{r}})  = \frac{\sigma_r^2}{N_s},
\end{equation}
with $\sigma_r$ being the root-mean-squared width of the beam spot (assuming circular symmetry). Note that the above variance does not take into account the effect of the pixel size, a more thorough analysis taking this into account can be found in \cite{Jia2010}.

Thus, the variance in measuring a displacement from position ${\bm{r}}_1$ to ${\bm{r}}_2$ is given by:
\begin{equation}
    \text{Var}(\Delta\bar{\bm{r}}) = \text{Var}(\bar{\bm{r}}_1) + \text{Var}(\bar{\bm{r}}_2)= \frac{2}{N_s}\sigma_r^2,
    \label{variance}
\end{equation}
where we have assumed that the beam width and total number of photon detections for estimating ${\langle \vec{\bm{r}}_1 \rangle}$ and ${\langle \vec{\bm{r}}_2 \rangle}$, are approximately equal under the same data acquisition time. Equation~\eqref{variance} also applies to the Fourier plane, which gives the variance in the momentum, providing directional information.

Limited by the HUL, the product of the beam width in the position $\sigma_r$ and momentum plane $\sigma_k$ is
\begin{equation}
    \sigma_r (\hbar \sigma_{k}) \geq \hbar/2.
\end{equation}
As such, the uncertainty product in the change of position and momentum have the following bound
\begin{equation}
    \sqrt{\text{Var}(\Delta\bar{\bm{r}})\text{Var}(\Delta\bar{\bm{k}})} = \frac{2}{N_s}\sigma_r \sigma_k\geq \frac{1}{N_s}.
    \label{HUL}
\end{equation}

Now we will show that by making joint measurements on a pairs of position-momentum entangled photons, it is possible to achieve an uncertainty product in the change of position and momentum with a precision exceeding this bound.

In the low gain regime, the position-momentum entangled state of SPDC in transverse momentum space can be written as
\begin{equation}
    |\Psi\rangle = \int\int \phi(\bm{k}_s,\bm{k}_i) |\bm{k}_s,\bm{k}_i\rangle\, d^2k_s d^2k_i,
\end{equation}
and the biphoton wavefunction $\phi(\bm{k}_s,\bm{k}_i)$ under a double-Gaussian approximation is given by \cite{Law2004,Chan2007,Schneeloch2016}

\begin{align}
    \phi({\bm{k}}_s,{\bm{k}}_i)\approx & A_k\exp\left(\frac{-\delta_r^2|\bm{k}_s-\bm{k}_i|^2}{8}\right)\nonumber\\
    &\times\exp\left(\frac{-|\bm{k}_s+\bm{k}_i|^2}{2\delta_k^2}\right),
    \label{kcorr}
\end{align}
where $A_k$ is a normalization constant, $\delta_k \approx 1/(2\sigma_p)$ with $\sigma_p$ being the pump beam width and $\delta_r\approx\sqrt{\frac{2\alpha L \lambda_p}{\pi}}$ where $L$ is the crystal length, $\lambda_p$ is the pump wavelength, and $\alpha = 0.455$ is a constant factor from the Gaussian approximation of the sinc phase matching function~\cite{Chan2007}.

Inverse Fourier transforming to position space, the biphoton wavefunction is then
\begin{align}
    \psi({\bm{r}}_s,{\bm{r}}_i)\approx & A_r\exp\left(\frac{-|\bm{r}_s-\bm{r}_i|^2}{2\delta_r^2}\right)\nonumber\\
    &\times\exp\left(-2\delta_k^2|\bm{r}_s+\bm{r}_i|^2\right).
    \label{rcorr}
\end{align}
From Eqs.~\eqref{kcorr} and \eqref{rcorr}, we can see that the width of position correlation $\bm{r}_s-\bm{r}_i$ is $\delta_r$ and for the transverse momentum correlation $\bm{k}_s+\bm{k}_i$ width is $\delta_k$. Utilizing these correlation measurements allows us to express the variance in the relative position and momentum as:
\begin{align}
      \text{Var}(\Delta\bar{\bm{r}}) =  &\frac{\Delta|\bm{r}_s - \bm{r}_i|}{\sqrt{N_c}} = \sqrt{\frac{2}{N_c}}\delta_r,\nonumber\\
     \text{Var}(\Delta\bar{\bm{k}}) =  &\frac{\Delta|\bm{k}_s + \bm{k}_i|}{\sqrt{N_c}} = \sqrt{\frac{2}{N_c}}\delta_k,
    \label{varianceSPDC}
\end{align}
where $N_c$ is the number of detected photon pairs. The position-momentum uncertainty relation in Eq.~\eqref{HUL} then becomes
\begin{equation}
    \sqrt{\text{Var}(\Delta\bar{\bm{r}})\text{Var}(\Delta\bar{\bm{k}})} = \frac{2}{N_c}\delta_r\delta_k
\label{HULSPDC}
\end{equation}

Thus, if the product $\delta_r\delta_k<1/2$, then the precision in measuring a change in beam trajectory can exceed the HUL.

\begin{figure}
    \centering
    \includegraphics[width=1\linewidth]{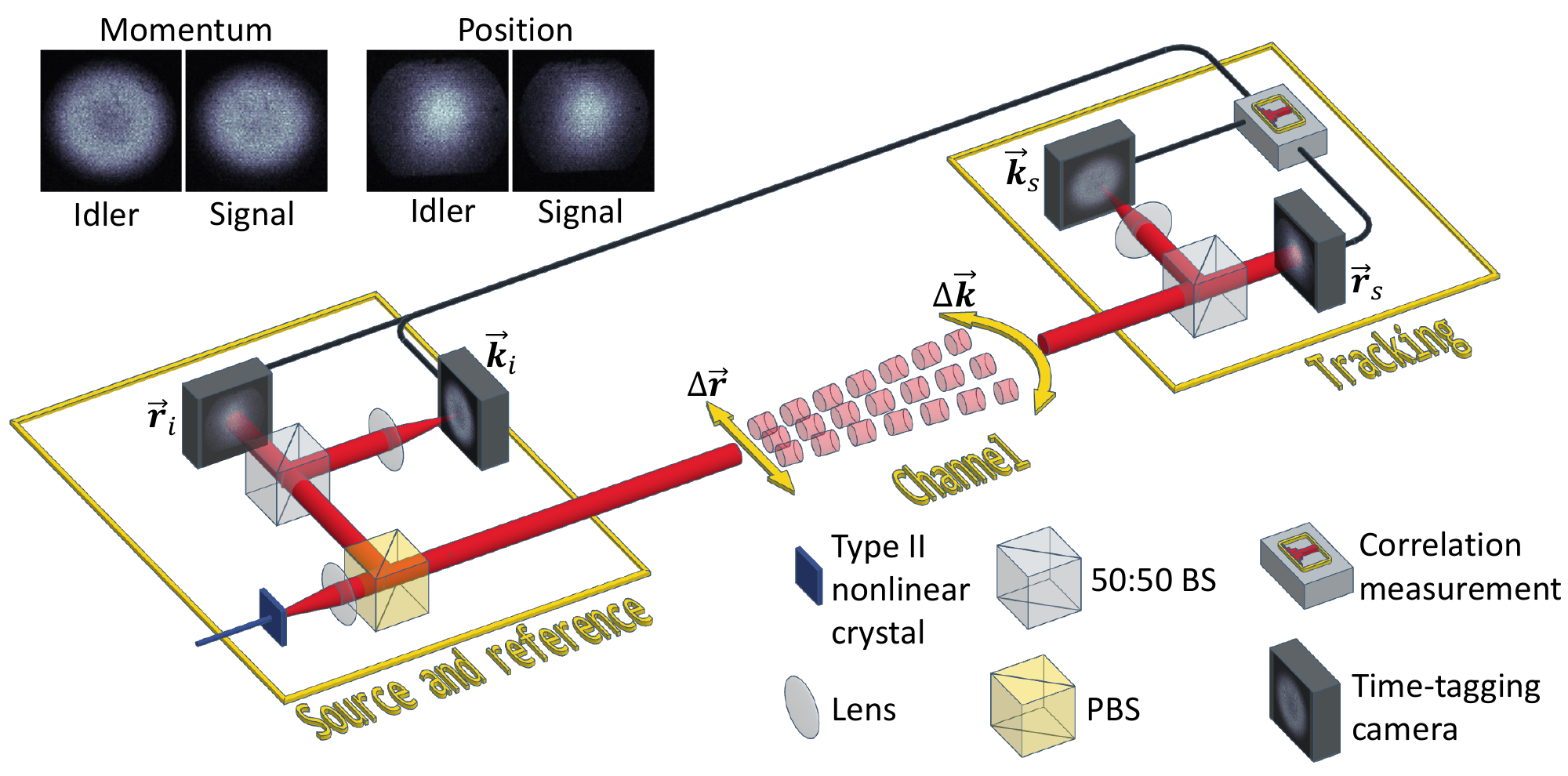}
    \caption{\textbf{Conceptual setup for quantum correlation beam tracking:} In the `Source and reference', position-momentum entangled photon pairs with orthogonal polarization are created through Type-II SPDC. One photon of the pair is split off using a PBS and then probabilistically split into being locally detected either in the position-plane or the momentum-plane of the non-linear crystal by time-tagging cameras. The other photon is sent through a `Channel' whose stability needs to be monitored. In the `Tracking' part of the setup, the photon is also probabilistically split into being detected in either the near or far field of the non-linear crystal by time-tagging cameras. A change in the trajectory of the signal photon beam is then tracked through displacements in the position and momentum correlation between the two photons. Typical images captured by the camera of the two planes for the photon pairs are shown on the top left.}
    \label{Fig1}
\end{figure}
A conceptual sketch of the setup used to realize QCBT is shown in Fig.~\ref{Fig1}. Position-momentum entangled photon pairs, known as the signal and idler photons, with orthogonal polarization are generated through Type-II SPDC. The photon pairs are then deterministically separated using a polarizing beam splitter (PBS) with the idler photons kept locally in a well-isolated system as a reference, while the signal photons are sent through a channel, which creates an unknown, time-varying, displacement on the beam. For detection, both photons are probabilistically split using a balanced beam splitter such that either the crystal's position-plane (near-field) or the momentum-plane (far-field) is measured by time-tagging cameras, which can capture both the spatial and time-of-arrival information of the photons. 

First, a time correlation measurement is performed between the two position-plane cameras and two momentum-plane cameras to identify photons that are detected in coincidence. Thereafter, a spatial correlation measurement is performed, whereby $\bm{r}_s-\bm{r}_i$ is determined from the photon pairs captured in the position-planes and $\bm{k}_s+\bm{k}_i$ determined through pairs from the momentum-planes.

For this proof-of-concept demonstration, we used a single camera (TPX3CAM~\cite{Nomerotski2019,ASI2}) to capture both the signal and idler photons in the position and momentum plane instead of measuring them separately with multiple cameras. \COR{We emulate the action of an arbitrary channel displacement by impinging the signal beam on a translating mirror. Because the mirror translates in a direction that is not parallel to the incoming beam, it introduces a displacement in both the angle and position of the beam.} Results from using the spatial correlations of SPDC photon pairs are compared to a coherent laser where the beam is tracked through a single photon on the signal beam arm. More details on the experimental setup can be found in the Supplementary Materials.

\begin{figure}
    \centering
    \includegraphics[width=1\linewidth]{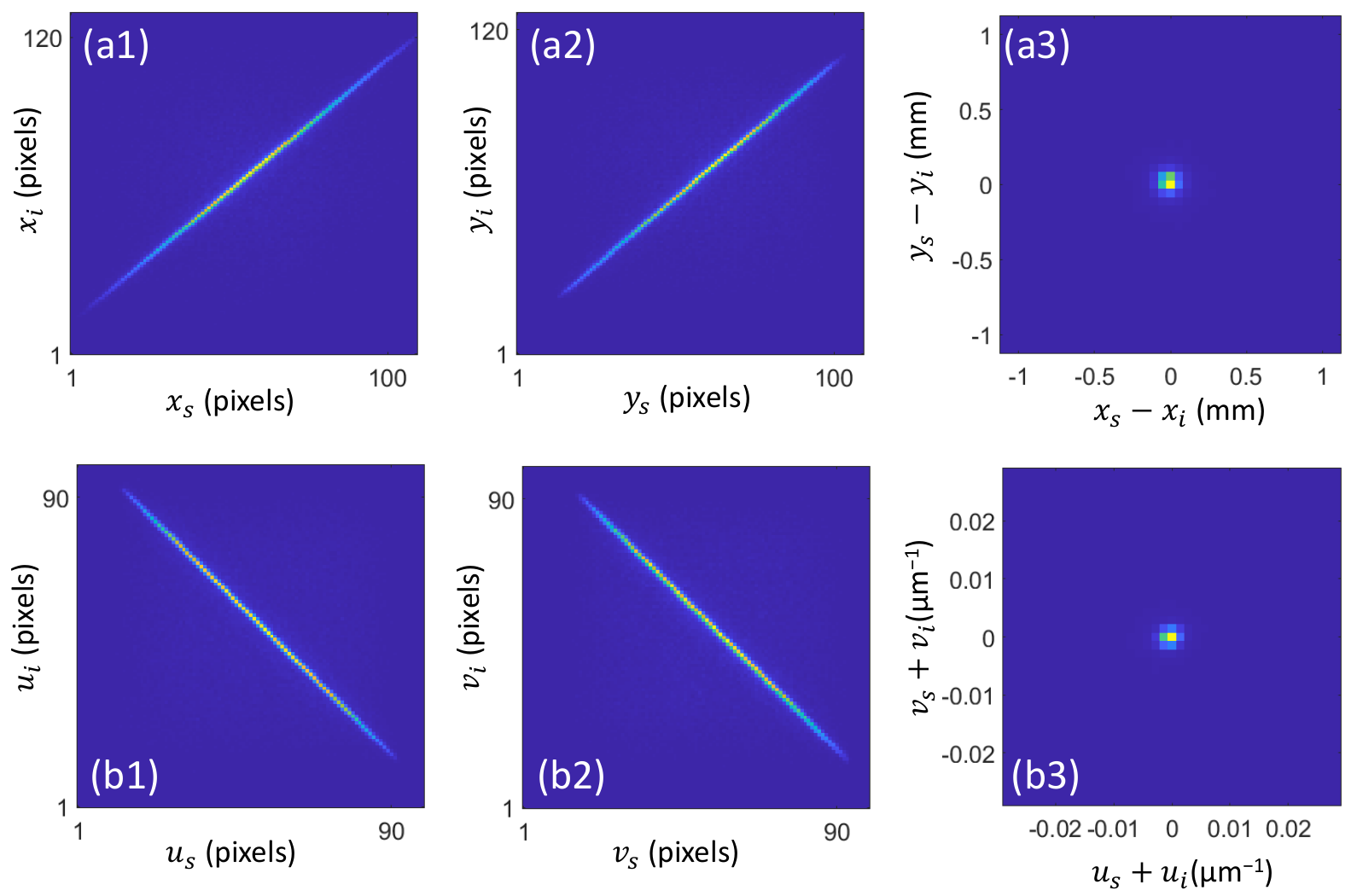}
    \caption{\textbf{SPDC correlation in near- and far-field:} (a1, a2) Measured position correlation of SPDC photons in the x and y direction. The difference-coordinate projection of position correlation is shown in (a3). (b1, b2) Measured momentum anti-correlation of the SPDC photons in the $u$ and $v$ direction. The sum-coordinate projection of momentum anti-correlation is shown in (b3).}
    \label{Fig2}
\end{figure}

The measured spatial correlations of SPDC at the camera plane is shown in Fig.\,\ref{Fig2}. When a Gaussian is fitted, a width of $\delta_r = 42\pm2$\,$\mu$m (0.76 pixels) is obtained for the $\bm{r}_s-\bm{r}_i$ profile, and $\delta_k = (1.06\pm0.04)\times10^{-3}$\,$\mu$m$^{-1}$ (0.74 pixels) for the $\bm{k}_s+\bm{k}_i$ profile. This would give a position-momentum uncertainty product of $(0.088\pm0.005)/N$, indicating that a measurement precision beyond the HUL should be possible. For the laser beam we have $\sigma_r=52.7\pm 0.7$\,$\mu$m and $\sigma_k=(1.41\pm0.02)\times10^{-2}$\,$\mu$m$^{-1}$, giving an uncertainty product of $(1.49\pm0.03)/N$, thus not surpassing the HUL.

\begin{figure}
    \centering
    \includegraphics[width=1\linewidth]{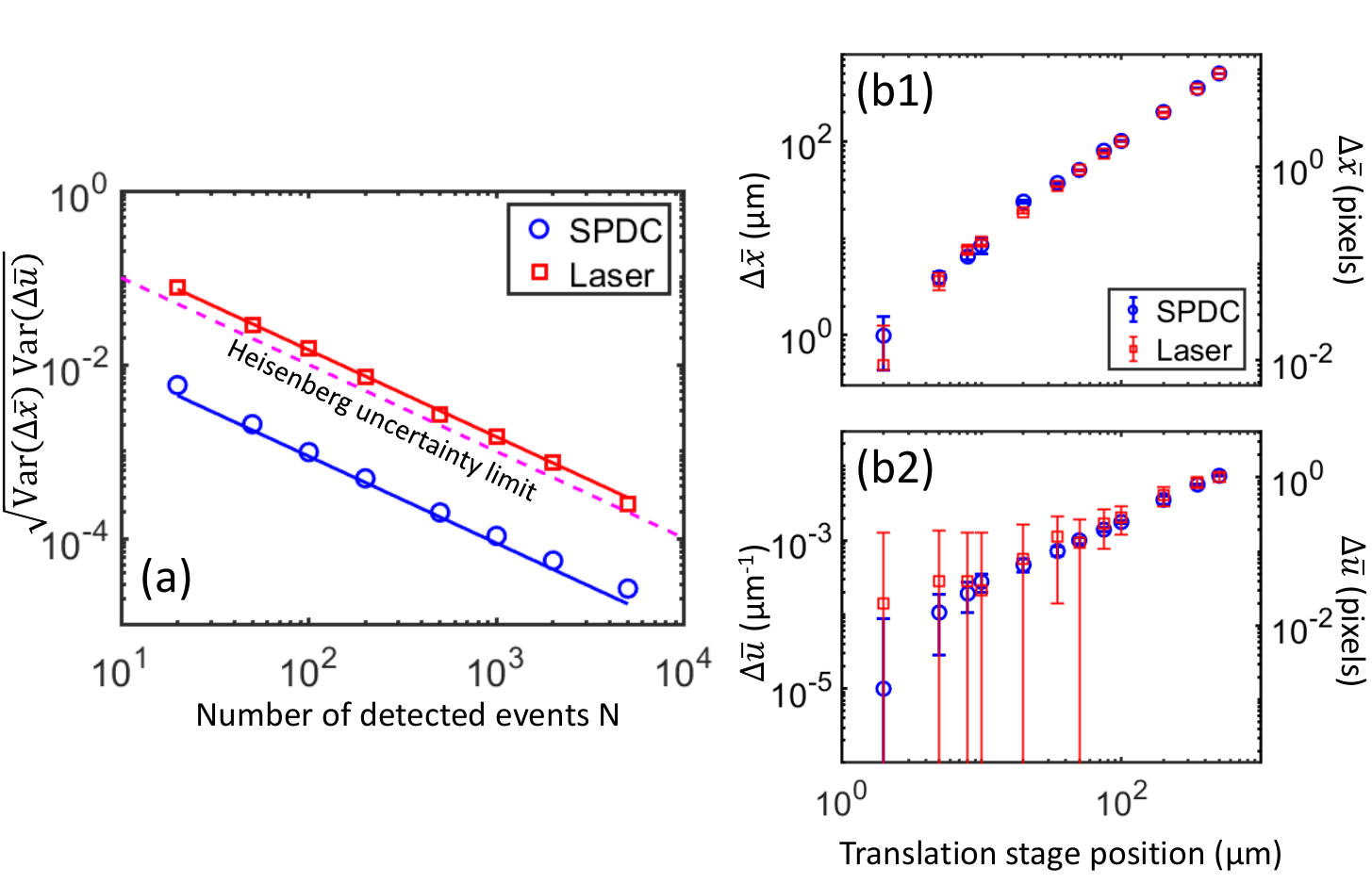}
    \caption{\textbf{Beam tracking accuracy:} (a) Position-momentum uncertainty product for the change in beam trajectory as a function of the number of detected events. This is compared for the cases of using the spatial correlations of SPDC (blue circles) and a laser (red squares). The solid lines are the expected uncertainty product based on the measured correlation and beam widths. The purple dashed line is the HUL corresponding to $\sqrt{\text{Var}(\Delta\bar{\bm{r}})\text{Var}(\Delta\bar{\bm{k}})} = 1/N$. (b1, b2) compares the measured beam displacement in the position and momentum planes for $N=5000$ while using SPDC spatial correlations (blue circles) and a laser (red squares).}
    \label{Fig3}
\end{figure}

Figure~\ref{Fig3} shows the experimental results of beam tracking with the two light sources. Figure~\ref{Fig3}(a) shows the uncertainty product for the two sources as a function of the number of detected events $N$. The average uncertainty product, determined through 50 repeated measurements at each $N$, are $(0.114\pm0.005)/N$ and $(1.54\pm0.04)/N$ for the correlated SPDC and laser, respectively. 


In Fig.~\ref{Fig3}(b) we translate a mirror by various distances and compare the measured displacement in the position (Fig.~\ref{Fig3}(b1)) and momentum (Fig.~\ref{Fig3}(b2)) plane for the laser and the correlated SPDC. This is measured at $N=5000$ with uncertainty obtained through 50 repeated measurements. Note that the mirror translates the beam horizontally, so position (momentum) shifts are only measured in the $x$ ($u$) direction. In the position plane ($\Delta x$), SPDC achieves comparable uncertainty to the laser since $\delta_r\simeq \sigma_r$, However, in the momentum plane ($\Delta u$), the SPDC offers far smaller uncertainty since $\delta_k\ll \sigma_k$: this is a direct visualization of the advantage of using entangled light for beam tracking.

\COR{Now one might wonder whether overlapping two classical lasers — one with a narrow beam width in the position plane and the other narrow in the momentum plane — followed by separation via polarization or wavelength just before detection, could achieve a similar accuracy as quantum correlations. It is easy to show using ray-transfer matrices, that for the measured change in beam trajectory by each of the laser to be equal will require perfect overlap of the two lasers. However, the task of overlapping these two lasers is, in itself, constrained by the uncertainty principle. One cannot achieve a perfect overlap where the beam centroids in both the position and momentum planes coincide with an accuracy smaller than twice the HUL. Detailed calculations of the ray-transfer matrices and beam overlap measurement can be found in the supplementary materials.}

It is important to note that a detection event is defined differently for different light sources. For the laser, this is the detection of a single photon on one of the tracking cameras, so $N=N_s$. For the correlated SPDC measurement, this is the coincident detection of a signal and an idler photon, thus $N=N_c$. For realistic detectors, $N_s$ and $N_c$ are related by $N_c = \epsilon_iN_s$ where $\epsilon_i\leq1$ is the detection efficiency of the idler photons. In this case, the HUL bound of Eq.~\eqref{HULSPDC} is modified into 
\begin{equation}
   \sqrt{\text{Var}(\Delta\bar{\bm{r}})\text{Var}(\Delta\bar{\bm{k}})} = \frac{2\delta_r\delta_k}{\epsilon_iN_s}.
\end{equation}
Thus, unconditionally beating the HUL will require $\epsilon_i > 2\delta_r\delta_k$, which for our demonstration $\epsilon_i$ must be greater than 0.11. Unfortunately for this experimental system, $\epsilon_i \approx 0.02$, which takes into account the camera detection efficiency of $\sim 0.08$ \cite{Vidyapin2022}, losses in the system, and statistically splitting the photons between the position and momentum planes which accounts for a 50$\%$ loss in the spatial-temporal correlated events. 

In addition to using a higher efficiency camera to improve $\epsilon_i$, one can try to circumvent the need to statistically split the photons by modifying the time-tagging camera into a light-field camera~\cite{Xiao2013,Martinez2018}, enabling simultaneous measurement of position and momentum information of the photons. One can also design the SPDC source to have smaller $\delta_r$ and $\delta_k$ to reduce the required $\epsilon_i$. However, both approaches require that the time-tagging camera have a higher spatial resolution than that available here. With the rapid advance of single-photon camera technologies~\cite{ASI1,Canon}, we expect that a time-tagging camera with enough efficiency and resolution to unconditionally beat the HUL will be available in the near future.

We also demonstrate that it is now possible to perform real-time measurements of the temporal-spatial correlations of SPDC and use this for beam tracking through QCBT. Using an Intel Xeon W-2145 CPU, we can process the temporal and spatial correlation of around $10^5$ detection events per second resulting in $\sim 100$ correlated events in both the position and momentum planes. This allows tracking of the beam displacement with an accuracy of $~\sim 5$\,$\mu$m (0.1 pixels). This is shown in the Supplementary Video.  With the latest generation of CPUs or possibly with GPU processing, we expect the speed of data processing can be significantly improved, allowing for faster tracking speed and higher accuracy. Details on TPX3CAM data processing can be found in \cite{Zhao2017,Vidyapin2022}. 

\begin{figure}[h]
    \centering
    \includegraphics[width=\linewidth]{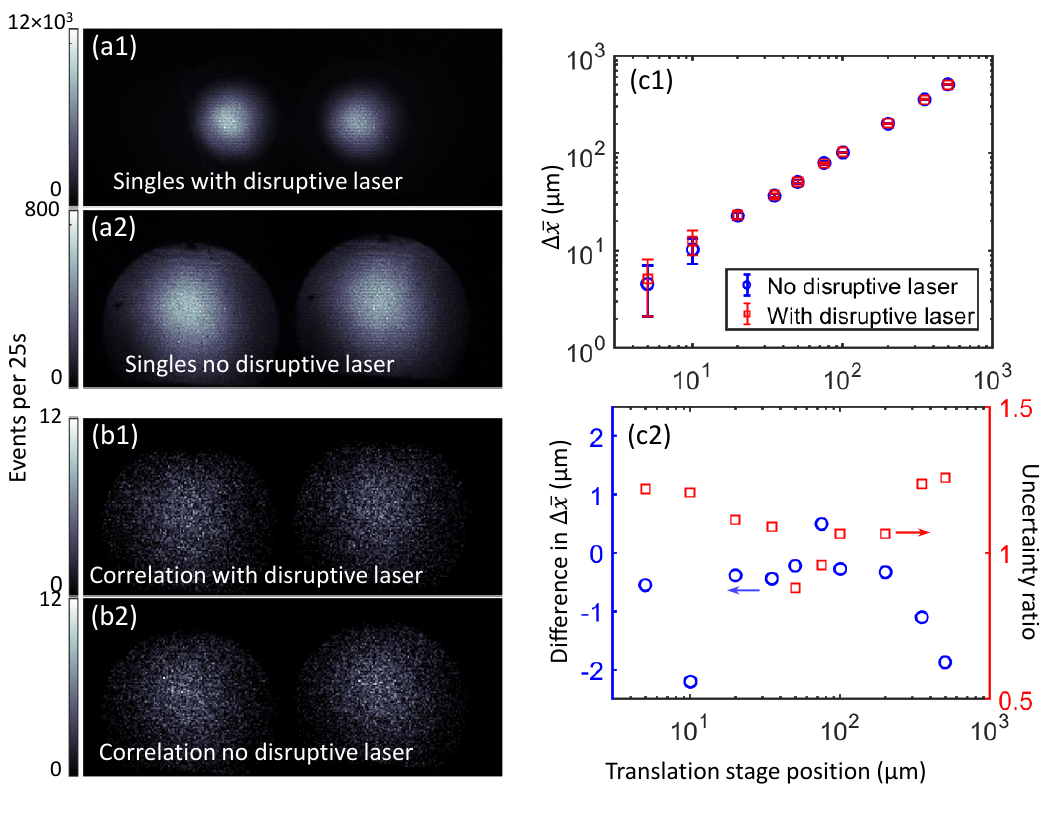}
    \caption{\textbf{Beam tracking accuracy under a strong background:} (a1, a2) Image from just singles events of the SPDC light with and without a bright disruption laser applied. (b1, b2) Detected temporally and spatially correlated photons with and without the disruption laser being applied. (c1, c2) Comparing the measured beam displacement from 1000 correlated events for QCBT with and without the disruption laser. (b2) shows the difference in $\Delta\bar{x}$ and the ratio of the uncertainties measured with and without the disruption laser. Uncertainties are determined through 50 repeated measurements}
    \label{Fig4}
\end{figure}

Beam tracking can be very sensitive to background influences; any incomplete background removal will lead to an incorrect centroid measurement. \COR{As we show in the supplementary, with a flat background of just $1/100$ the laser brightness can increase the measured centroid uncertainty by a factor of 2.5.}  QCBT exibits extreme background tolerance through the time and spatial correlation of SPDC photons. Here, a disruptive laser beam that has approximately 10 times more photons than SPDC is applied on both the signal and idler regions of the camera, as seen in Fig,~\ref{Fig4}(a). After post-selecting on temporally and spatially correlated photons, we can see in Fig,~\ref{Fig4}(b) that the disruptive laser beam is completely removed. \COR{The result of beam tracking in the position plane comparing the cases of having the disruptive laser beam on or off is shown in Fig,~\ref{Fig4}(c). The two scenarios show near-perfect agreement in the measured beam displacement. For 1000 correlated events, $\Delta\bar{x}$ differed by no more than 1 micron on average, and the uncertainty only increased by an average of 10 percent.}

In conclusion, we have demonstrated a proof-of-principle quantum enhanced beam tracking technique using position-momentum entangled SPDC photons and have shown how it can be used to achieve a beam tracking accuracy beyond the bound of the HUL. Assuming perfect detectors, QCBT with our system can achieve a position-momentum uncertainty product of $0.11/N$ compared to the HUL bound of $1/N$. We also demonstrate real-time beam tracking using QCBT at a speed of $\sim1$\,Hz with an accuracy of 5\,$\mu$m through just 100 correlated photons. QCBT is also extremely resistant to background influences; we show that even when a disruptive laser that is ten times brighter than SPDC is applied, there is still negligible effect on the accuracy of QCBT. \COR{With the combination of violating the HUL and strong noise resilience, QCBT can be desirable for applications requiring low-transmitted power such as sensing of light-sensitive samples, covert applications or stabilizing quantum communication channels.}

\bibliography{BTref}

\vspace{1cm}
\noindent\textbf{Acknowledgment:} The authors are grateful to Aaron Goldberg for stimulating discussion. This work was supported by NRC-uOttawa Joint Centre for Extreme Quantum Photonics (JCEP) via the Quantum Sensors Challenge Program at the National Research Council of Canada, Quantum Enhanced Sensing and Imaging (QuEnSI) Alliance Consortia Quantum grant, and the Canada Research Chair (CRC) Program.

\appendix
\section{Details on the experimental setup}

\begin{figure*}
    \centering
    \includegraphics[width=0.85\linewidth]{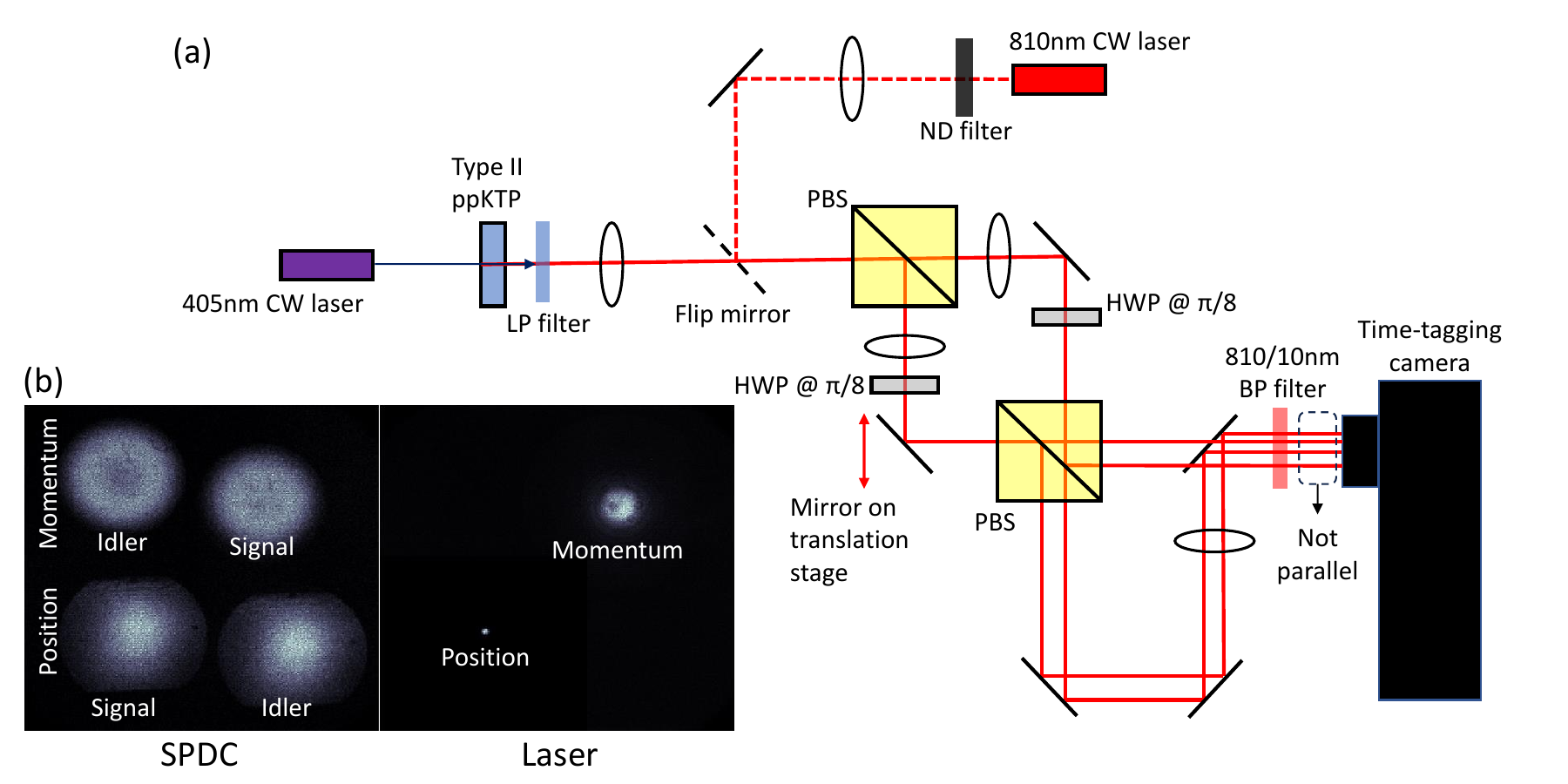}
    \caption{\textbf{Experimental setup for SPDC beam tracking:} (a) Experimental setup for demonstrating QCBT. LP-filter: long-pass filter, BP-filter: band-pass filter, ND-filter: neutral density filter, PBS: polarizing beam-splitter, HWP: half-wave plate (b) Image captured on camera of the position and momentum planes from SPDC and laser. Note that in order to also see the laser momentum plane in the same image, the laser position plane intensity is displayed at 1/30 of the actual intensity.}
    \label{Setup}
\end{figure*}

The experimental setup demonstrating QCBT is shown in Fig.~\ref{Setup}(a). Position-momentum entangled photon pairs with orthogonal polarization are generated by pumping a 1\,mm thick Type-II ppKTP crystal with a 405\,nm CW laser. The two photons are first separated using a PBS and then, using HWPs, the polarization of each photon is rotated by $\pi/4$ so at the second PBS each photon will have a 50\% probability of being imaged either at the position-plane of the crystal or the momentum-plane of the crystal. To create a beam displacement, a mirror mounted on a motorized translation stage is placed in between the crystal's position and momentum planes. The photon trajectories are slightly misaligned after the second PBS such that their trajectories are not perfectly parallel to each other; this way, moving the mirror will cause a simultaneous displacement of the beam in both the position and momentum planes. 

For comparison of QCBT with the best classical scenario, the setup can be switched between using a SPDC source or an attenuated 810\,nm laser by a flip mirror. The time-tagging camera used for detection is the TPX3CAM which has a resolution of 256$\times$256 pixels with a pixel pitch of 55\,$\mu$m and is able to time tag the photons detected on each pixel with $\sim 7$\,ns accuracy. Typical images captured with the camera for the position and momentum planes of SPDC and laser are shown in Fig.~\ref{Setup}(b).

The expected correlation widths can be approximated according to\,\cite{Chan2007}:
\begin{align}
    \delta_k &\approx 1/(2\sigma_p), \nonumber \\
    \delta_r &\approx\sqrt{\frac{2\alpha L \lambda_p}{\pi}}
\end{align}
With the experimental parameters $L=1$\,mm, $\lambda_p = 405$\,nm, $\sigma_p = 0.12$\,mm and a $5\times$ magnification in the imaging system from the crystal to the camera, we expect $\delta_r = 54$\,$\mu$m and $\delta_k = 0.83\times10^{-3}$\,$\mu$m$^{-1}$ at the camera. These values are in approximate agreement with the measured values of $\delta_r = 42$\,$\mu$m and $\delta_k = 1.1\times10^{-3}$\,$\mu$m$^{-1}$.

\section{Optimal beam centroid estimator}
To show that the center of mass equation 
\begin{equation}
    \bar{x} = \frac{\sum_{j=1}^N x_j}{N},
\end{equation}
is the most optimal estimator of the beam centroid $x_c$, we first note that $\bar{x}$ is an unbiased estimator of $x_c$ since the expectation value of $\bar{x}$ 
\begin{align}
    E[\bar{x}] &= E\left[\frac{\sum_{j=1}^N x_j}{N}\right]\nonumber\\
    &= \frac{\sum_{j=1}^N E[x_j]}{N}\nonumber\\
    &= x_c.
\end{align}

Secondly, we must show that $\bar{x}$ saturates the Cramer-Rao bound
\begin{equation}
    \text{Var}(\bar{x}) = \frac{1}{NI},
\end{equation}
where $I$ is the Fisher information given by
\begin{equation}
    I = E\left[\left(\frac{\partial}{\partial x_c}\log P(x|x_c,\sigma)\right)^2 \right],
\end{equation}
with $P(x|x_c,\sigma)$ the probability to detect a photon at position $x$ given the beam's centroid $x_c$ and a root-mean-squared waist $\sigma = \sqrt{\frac{\sum_{j=1}^N (x_j - \bar{x})^2}{N}} = \sqrt{\text{Var}\left(x_j\right)}$. 

The variance $\text{Var}(\bar{x})$ is 
\begin{align}
    \text{Var}(\bar{x}) &= \text{Var}\left(\frac{\sum_{j=1}^N x_j}{N}\right) \nonumber\\
    &= \frac{1}{N^2}\text{Var}\left(\sum_{j=1}^N x_j\right) \nonumber\\
    &= \frac{1}{N^2}\sum_{j=1}^N \text{Var}\left(x_j\right) \nonumber\\
    &= \frac{\sigma^2}{N},
\label{var}
\end{align}
where in the 3rd line we have assumed that the photon positions $x_i$ are independent.

Assuming a Gaussian distribution for $P(x|x_c,\sigma) = \frac{1}{\sqrt{2\pi \sigma^2}}\exp\left(\frac{-(x-x_c)^2}{2\sigma^2}\right)$,
\begin{align}
    I &=  E\left[\left(\frac{\partial}{\partial x_c} \frac{-(x-x_c)^2}{2\sigma^2}\right)^2 \right] \nonumber\\
    & = E\left[\frac{(x-x_c)^2}{\sigma^4} \right]\nonumber\\
    & = \frac{1}{\sigma^2},
\end{align}
thus $1/NI = \sigma^2/N$, which is the variance given by Eq.\ref{var}. This signifies that, when assuming a Gaussian probability distribution, the center of mass equation does satisfy the Cramer-Rao bound and is thus the most optimal beam centroid estimator. 

\section{Uncertainty from overlapping two lasers}

\begin{figure*}
    \centering
    \includegraphics[width=0.85\linewidth]{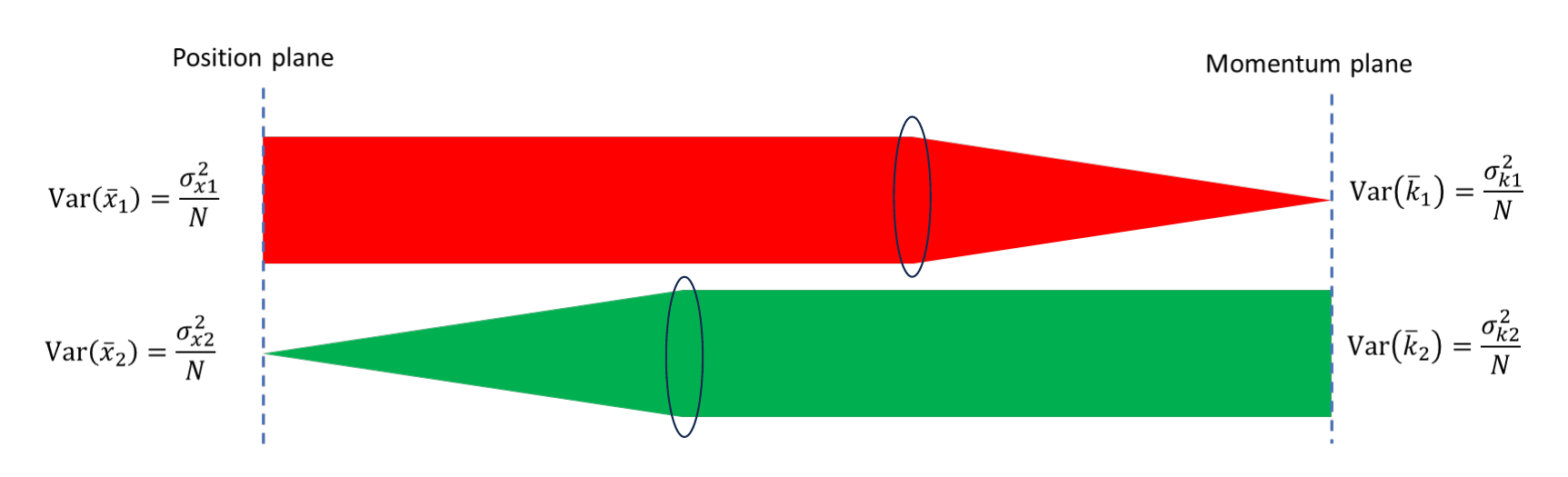}
    \caption{Overlapping two laser beams of different wavelength}
    \label{SuppFig2}
\end{figure*}

We want to overlap the two laser beams as shown in Fig.\,\ref{SuppFig2} where the variance in the beam centroid is also shown for each beam and plane. Here, $\sigma_{x1(2)}$ and $\sigma_{k1(2)}$ are the beam width in the position and momentum plane respectively for beam 1(2). For simplicity we will assume the number of photons $N$ is the same in each beam and plane. The variance in the centroid from overlapping the two beams is then
\begin{align}
    \text{Var}(\bar{x}_\text{overlap}) &= \frac{\sigma_{x1}^2}{N} +  \frac{\sigma_{x2}^2}{N}\nonumber\\
    \text{Var}(\bar{K}_\text{overlap}) &= \frac{\sigma_{k1}^2}{N} +  \frac{\sigma_{k2}^2}{N},
\end{align}
and the variance for a change in centroid when the beam is moved (assuming no change to the beam width) is 
\begin{align}
    \text{Var}(\Delta\bar{x}_\text{overlap}) &= 2\text{Var}(\bar{x}_\text{overlap})\nonumber\\
    \text{Var}(\Delta\bar{k}_\text{overlap}) &= 2\text{Var}(\bar{k}_\text{overlap}).
\end{align}
The resulting uncertainty product is then
\begin{align}
    &\sqrt{\text{Var}(\Delta\bar{x}_\text{overlap})\text{Var}(\Delta\bar{k}_\text{overlap})} \nonumber\\
     =&\frac{2}{N}\sqrt{\sigma_{x1}^2\sigma_{k1}^2 + \sigma_{x2}^2\sigma_{k2}^2 + \sigma_{x1}^2\sigma_{k2}^2 + \sigma_{x2}^2\sigma_{k1}^2}.
\end{align}
Given the HUL $\sigma_{x1(2)}\sigma_{k1(2)}\geq 1/2$, therefore
\begin{align}
    &\sqrt{\text{Var}(\Delta\bar{x}_\text{overlap})\text{Var}(\Delta\bar{k}_\text{overlap})} \nonumber\\
    \geq& \frac{1}{N}\sqrt{1+1+\alpha^2+\frac{1}{\alpha^2}}\nonumber\\
    \geq& \frac{2}{N},
\end{align}
where $\alpha = \frac{\sigma_{x1}}{\sigma_{x2}} = \frac{\sigma_{k2}}{\sigma_{k1}}$. 

We can see that the uncertainty product for overlapping the two beams is at least twice that for a single beam, as given by Eq.\,5 of the main text, and this occurs for $\alpha = 1$ in which the two beams have the same beam width in both planes. For any other values of $\alpha$, the uncertainty product becomes even larger.

The calculation would be similar for any other scenarios where one would want to match beam trajectories, such as quickly switching between two lasers where each laser is focused in a different plane, or switching between two lens configurations, which has the laser focused at different planes for each configuration.

One would also question the importance of perfect beam overlap if one only needs to measure a change in the beam trajectory. Below we show through a simple example using ray-transfer matrices that in general the change in beam trajectory will be different if the two beams are not perfectly overlapped. Assume the following scenario where we have a channel of length d sandwiched between two imaging lenses of focal length f (which is how the experiment is setup) with the initial position and angle of the beam being $\begin{pmatrix}
r_i\\ 
\theta_i
\end{pmatrix}$:
\begin{figure}[h]
    \centering
    \includegraphics[width=1\linewidth]{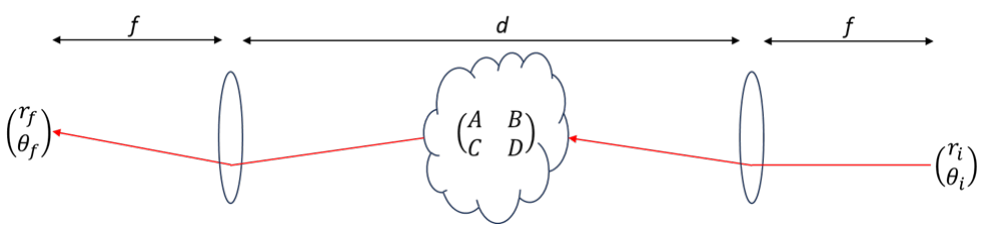}
    \label{SuppFig3}
\end{figure}

Now assume initially the channel does not change the beam trajectory so the ray transfer matrix of the channel is just $\begin{pmatrix}
1 & d\\ 
0 & 1
\end{pmatrix}$ so the final position and angle of the beam is:
\begin{widetext}
\begin{equation}
\begin{pmatrix}
r_f\\ 
\theta_f
\end{pmatrix} 
= 
\begin{pmatrix}
1 & f\\ 
0 & 1
\end{pmatrix}
\begin{pmatrix}
1 & 0\\ 
-1/f & 1
\end{pmatrix}
\begin{pmatrix}
1 & d\\ 
0 & 1
\end{pmatrix}
\begin{pmatrix}
1 & 0\\ 
-1/f & 1
\end{pmatrix}
\begin{pmatrix}
1 & f\\ 
0 & 1
\end{pmatrix}
\begin{pmatrix}
r_i\\ 
\theta_i
\end{pmatrix}
= \begin{pmatrix}
-r_i\\ 
\frac{-2f+d}{f^2}r_i-\theta_i
\end{pmatrix}.
\end{equation}
\end{widetext}
Now if the channel makes a random change to the beam trajectory with ray transfer matrix $\begin{pmatrix}
A & B\\ 
C & D
\end{pmatrix}$ then the final position and angle of the beam becomes:
\begin{widetext}
\begin{align}
\begin{pmatrix}
r_f\\ 
\theta_f
\end{pmatrix} 
&= 
\begin{pmatrix}
1 & f\\ 
0 & 1
\end{pmatrix}
\begin{pmatrix}
1 & 0\\ 
-1/f & 1
\end{pmatrix}
\begin{pmatrix}
A & B\\ 
C & D
\end{pmatrix}
\begin{pmatrix}
1 & 0\\ 
-1/f & 1
\end{pmatrix}
\begin{pmatrix}
1 & f\\ 
0 & 1
\end{pmatrix}
\begin{pmatrix}
r_i\\ 
\theta_i
\end{pmatrix}\nonumber\\
&= \begin{pmatrix}
(Cf-D)r_i + Cf^2\theta_i\\ 
\frac{[B-(A+D)f+Cf^2]r_i-[Af^2+Cf^3]\theta_i}{f^2}
\end{pmatrix}.
\end{align}
\end{widetext}

The resulting change in beam trajectory is thus
\begin{equation}
\begin{pmatrix}
\Delta r_f\\ 
\Delta \theta_f
\end{pmatrix} 
= \begin{pmatrix}
(Cf-D+1)r_i + Cf^2\theta_i\\ 
\frac{[B-d-(A+D-2)f+Cf^2]r_i-[(A+1)f^2+Cf^3]\theta_i}{f^2}
\end{pmatrix}.
\end{equation}
We see that the final change in position and angle depends on both the initial position and angle of the beam, so unless  $\begin{pmatrix}
r_i\\ 
\theta_i
\end{pmatrix}$ is the same for both beams, one will not measure the same change in beam trajectory.

\section{Influence of background on centroid measurement of laser}

\begin{figure}
    \centering
    \includegraphics[width=1\linewidth]{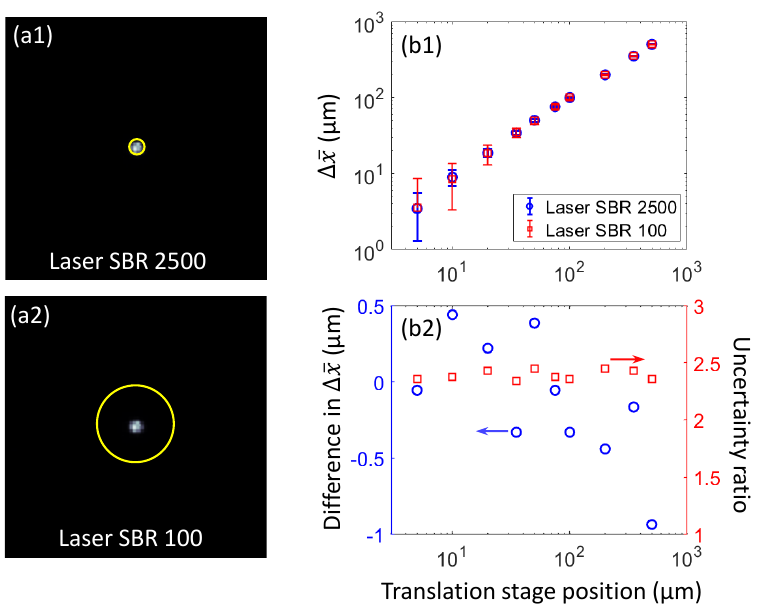}
    \caption{\textbf{Influence of a flat environmental background on the centroid measurement of a laser.} (a1, a2) shows the laser beam in the position plane with the yellow circles indicating the radius of the digital aperture used in post-processing to achieve a signal-to-background ratio (SBR) of 2500 and 100. (b1) shows the measured beam shift using 1000 photons in the position plane for the two SBR cases. (b2) shows the difference between the measured centroid shift and the ratio of the uncertainties. As can be seen, the difference in the measured centroid shift is less than a micron on average, but the average uncertainty for SBR of 100 is about 2.5 larger than that for a SBR of 2500. Errorbars in (b1) are determined through 250 repeated measurements.}
    \label{SuppFig3}
\end{figure}

A centroid measurement can be very sensitive to background influences. In this experiment there is a flat electronic and environmental background of around 1.7 photons per second per pixel, which needs to be accounted for when performing centroid measurements using a laser. We applied a digital aperture in post-processing, centered on the laser beam to limit the number of background photons that can contribute to the centroid measurement. Figure~\ref{SuppFig3} shows the effect on choosing the digital aperture radius and hence the background effect on the centroid measurement taken by a laser. Figure~\ref{SuppFig3}~(a1) shows the digital aperture used for the results of Fig.~3(b1) of the main text. The diameter of the digital aperture is 2.5 times of the laser beam FWHM and the signal-to-background ratio (SBR) of 2500. Figure\,\ref{SuppFig3}\,(a1) has the digital aperture diameter increased to 12.5 times of the beam FWHM and the SBR is reduced to 100. Figure~\ref{SuppFig3}~(b1, b2) compares the measurement accuracy between these two aperture sizes. We see that although the large aperture measured an average beam shift which agrees with that measured using a small aperture to within a micron, the average uncertainty increased by a factor of 2.5.

This small background is certainly not an issue when using SPDC correlations. We see in Fig.~4 of the main text, that the measurement accuracy of QCBT was minimally affected even when a disruptive beam with 10 times more photons than SPDC was applied. 

\end{document}